\documentclass[sigconf, natbib=true, anonymous=false]{acmart}

\AtBeginDocument{%
  }

\copyrightyear{2025}
\acmYear{2025}
\setcopyright{cc}
\setcctype{by}

\acmConference[ICTIR '25]{Proceedings of the 2025 International ACM SIGIR Conference on Innovative Concepts and Theories in Information Retrieval (ICTIR)}{July 18, 2025}{Padua, Italy}

\acmBooktitle{Proceedings of the 2025 International ACM SIGIR Conference on Innovative Concepts and Theories in Information Retrieval (ICTIR) (ICTIR '25), July 18, 2025, Padua, Italy}

\acmDOI{10.1145/3731120.3744603}
\acmISBN{979-8-4007-1861-8/2025/07}

\usepackage{longtable}
\usepackage{tikz}
\usepackage{balance}
\usepackage{multirow}
\usepackage{booktabs}
\usepackage{color}
\usepackage{arydshln}
\usetikzlibrary{positioning, shapes.multipart}

\begin{document}

\title{Probing Ranking LLMs: A Mechanistic Analysis for Information Retrieval}

\author{Tanya Chowdhury}
\affiliation{%
  \institution{University of Massachusetts Amherst}
    \state{MA}
  \country{USA}
}

\author{Atharva Nijasure}
\affiliation{%
  \institution{University of Massachusetts Amherst}
    \state{MA}
  \country{USA}
}

\author{James Allan}
\affiliation{%
  \institution{University of Massachusetts Amherst}
  \state{MA}
    \country{USA}
}

\begin{abstract}



Transformer networks, particularly those achieving performance comparable to GPT models, are well known for their robust feature extraction abilities. However, the nature of these extracted features and their alignment with human-engineered ones remain largely unexplored. In this work, we investigate the internal mechanisms of state-of-the-art, fine-tuned LLMs for passage reranking. We employ a probing-based analysis to examine neuron activations in ranking LLMs, identifying the presence of known human-engineered and semantic features. Our study spans a broad range of feature categories, including lexical signals, document structure, query-document interactions, and complex semantic representations, to uncover underlying patterns influencing ranking decisions.

Through experiments on four different ranking LLMs, we identify statistical IR features that are prominently encoded in LLM activations, as well as others that are notably missing. Furthermore, we analyze how these models respond to out-of-distribution queries and documents, revealing distinct generalization behaviors. By dissecting the latent representations within LLM activations, we provide an important initial step toward better intepretability and aim to improve both the interpretability and effectiveness of ranking models. Our findings offer crucial insights for developing more transparent and reliable retrieval systems. We release all necessary scripts and code to support further exploration.

\end{abstract}


\maketitle

\section{Introduction}
For many decades, the domain of passage retrieval and reranking has predominantly used algorithms grounded in statistical human-engineered features, typically derived from the query, the document set, or their interactions. In particular, the features in training and evaluation datasets such as the learning-to-rank MSLR dataset \cite{DBLP:journals/corr/QinL13} are largely derived from features manually developed to support successful statistical ranking systems. However, recent advances have seen state-of-the-art passage retrieval and ranking algorithms increasingly pivot to neural network-based models. The introduction of Large Language Models (LLMs) has notably widened the performance disparity between neural-based and traditional statistical rankers.

Access to open-source ranking models, particularly those exhibiting performance comparable to GPTs, provides a unique opportunity to explore the inner workings of transformer-based architectures. Neural networks are recognized for their robust feature extraction capabilities; however, the representation of these features remains elusive, making it challenging to discern their nature and potential correlation with features engineered by humans. As a result, despite their efficacy, neural networks present a challenge in terms of transparency: their feature representations are complex and the location of critical features within the network remains obscure. In this context, mechanistic interpretability \cite{rauker2023toward} aims to demystify the internal workings of transformer architectures, showing how LLMs process and learn information differently compared to traditional methods. Our research is motivated by the desire to determine whether the statistical features traditionally valued in algorithms like BM25 and tf*idf are somehow encoded within LLM architectures. This study stands to bridge the gap between neural and statistical methodologies, by using probes to generate hypotheses on the inner workings of the LLMs, offering insights that could enhance the field of information retrieval by enabling more intuitive explanations and better support for model design and analysis. 

\textbf{Summary:} This study is built upon four different LLM architectures: Llama2-7b, Llama2-13b, Llama3.1-8b, and Pythia-6.9b. For each of these we use or create a LoRa fine-tuned variant (e.g., RankLlama \cite{ma2023fine}), optimized for passage reranking tasks using the MS Marco dataset. These point-wise rankers demonstrate substantial accuracy improvements over their non-LLM counterparts. Our analysis concentrates on extracting activations from the MLP unit of each transformer block, which is posited to contain the key feature extractors \cite{geva2020transformer}. We aggregate these activations for each input sequence (query-document pairs) and assign labels to these pairs corresponding to each feature that is a target of our probe. Subsequently, we employ a regression with regularization to correlate these activations with the labels. Our findings reveal a pronounced representation of certain MSLR features within the ranking LLMs, while others are markedly absent. We also observe that distinct LLM  architectures capture very similar statistical IR features, when fine-tuned using the same dataset. These insights will allow  future researchers to formulate hypotheses concerning the underlying circuitry of ranking LLMs.

\textbf{Research Questions:}
In this study, we aim to explore several internal mechanistic aspects of ranking LLMs through probing techniques. Specifically, we seek to \emph{determine whether known statistical information retrieval (IR) features are present within the activations of LLMs}. We are also interested in \emph{identifying groups of features and understanding how they may combine or interact within the LLM's activations}. Additionally, we \emph{investigate whether LLMs contain components that mimic similarity scores from models like BERT or T5}. It is also of interest to \emph{find out if different LLM architectures encode the same statistical IR features within their activations}. Finally, this inquiry extends to examining whether the latent features encoded within activations of LLMs remain consistent or change when the model encounters \emph{out-of-distribution queries or documents}. By answering these questions, we aim to gain a deeper understanding of the inner workings of ranking LLMs and the extent to which they align with traditional IR methodologies.

\subsection{Contributions and Findings}
\setlength{\leftmargini}{1.5em}
\begin{enumerate}
\item We discovered that several human-engineered metrics, such as \textit{covered query term number}, \textit{covered query term ratio}, \textit{mean of stream length normalized term frequency}, and \textit{variance of tf*idf}, are prominently encoded  within LLM activations. In contrast, certain features, including \textit{sum of stream length normalized term frequency}, \textit{max of stream length normalized term frequency}, and \textit{BM25}, show no discernible representation in LLM activations.
\item We fine-tuned and probed two additional reranking models, Llama3.1-8b and Pythia-6.9b, resulting in RankLlama3-8b and RankPythia, alongside probing the existing RankLlama2-7b and RankLlama2-13b models. The results revealed broadly consistent outcomes across all models, suggesting that different LLM architectures, when fine-tuned similarly, encode latent features in a comparable way within their activations.
\item We found that the activation patterns of the listed features remain consistent even when the model encounters out-of-distribution queries or documents on RankLlama2-7b, Rank\-Llama3-8b and RankPythia-6.9b. However, they do not remain consistent for certain features of RankLlama2-13b, suggesting potential overfitting in the LLM during fine-tuning.
\item We identified specific combinations of MSLR features, like the sum of \emph{covered query term ratio, mean stream length normalized term frequency,} and \emph{variance of tf*idf}, along with the sum's square and cube, all of which also exhibit a strong correlation with LLM activations.
\item Finally, we show further evidence on the presence of these encoded features using Shapley values. All scripts, datasets, and fine-tuned ranking LLMs developed and/or used in this study are made publicly available.\footnote{\url{https://github.com/TaKneeAa/ProbingRankLlama}} 

\end{enumerate}

\section{Background \& Related Work}
We discuss concepts in Mechanistic Interpretability, in particular sparse probing. Then we touch on inner interpretability approaches in Information retrieval.

\subsection{Mechanistic Interpetability \& Sparse Probing} 
 Mechanistic Interpretability is the ambitious goal of gaining an algorithmic-level understanding of any deep neural network's computations \cite{rauker2023toward}. 
This goal can be achieved in numerous ways, some of them being by studying weights (e.g., weight masking \cite{csordas2020neural,wortsman2020supermasks}, continual learning \cite{de2019bias}), by studying individual neurons (e.g., excitement based \cite{zhou2014object}, activation patching~\cite{makelov2023subspace}, gradient-based \cite{ancona2019gradient}, perturbation and ablation based ~\cite{zhou2018revisiting}), by studying subnetworks (e.g., sparsity based ~\cite{meister2021sparse}, circuit analysis based ~\cite{fiacco2019deep}), by studying representations (e.g., tokens ~\cite{li2021implicit}, attention ~\cite{clark2019does}, probing ~\cite{belinkov2022probing,gurnee2023language,chen2023beyond}), by training sparse auto encoders ~\cite{makelov2024towards,kissane2024saes,cunningham2023sparse}, etc. In this study, we focus on studying the activations of individual neurons and also some groups of neurons with the help of probing regressors. To that end, we use concepts from a host of techniques listed above. 

While the concept of reverse-engineering specific neurons within large language models (LLMs) is relatively new, existing studies \cite{geva2020transformer, geva2022transformer} illustrate that the feed-forward layers of transformers, comprising two-thirds of the model's parameters, function as key-value memories. These layers activate in response to specific inputs, a mechanism we aim to demystify by reverse-engineering the activation function of these neurons. A notable challenge in this endeavor is the phenomenon of superposition, where early layers in LLMs select and store a vast array of features—often exceeding the number of available neurons—in a linear combination across multiple neurons. In contrast, later layers tend to focus on more abstract features, discarding those deemed non-essential \cite{gurnee2023finding}.

Probing aims to determine if a given representation effectively captures a specific type of information, as discussed by Belinkov~\cite{belinkov2022probing}. This technique employs transfer learning to test whether embeddings contain information pertinent to a target task. The three essential steps in probing include: (1) obtaining a dataset with examples that exhibit variation in a particular quality of interest, (2) embedding these examples, and (3) training a model on these embeddings to assess if it can learn the quality of interest. This method is versatile as it can utilize any inner representation from any model. However, a limitation of probing methods is that in some cases, a successful probe does not necessarily mean that the probed model actually utilizes that information about the data ~\cite{ravichander2020probing}. Belinkov \cite{belinkov2022probing} provides a comprehensive survey on probing methods for large language models, discussing their advantages, disadvantages, and complexities. Gurnee et al. ~\cite{gurnee2023finding} further introduce sparse probing, where they mine features of interest over groups of representations up to $k$ in size. We build upon sparse probing in this work. 

\begin{figure}
    \centering
    \includegraphics[width=0.48\textwidth]{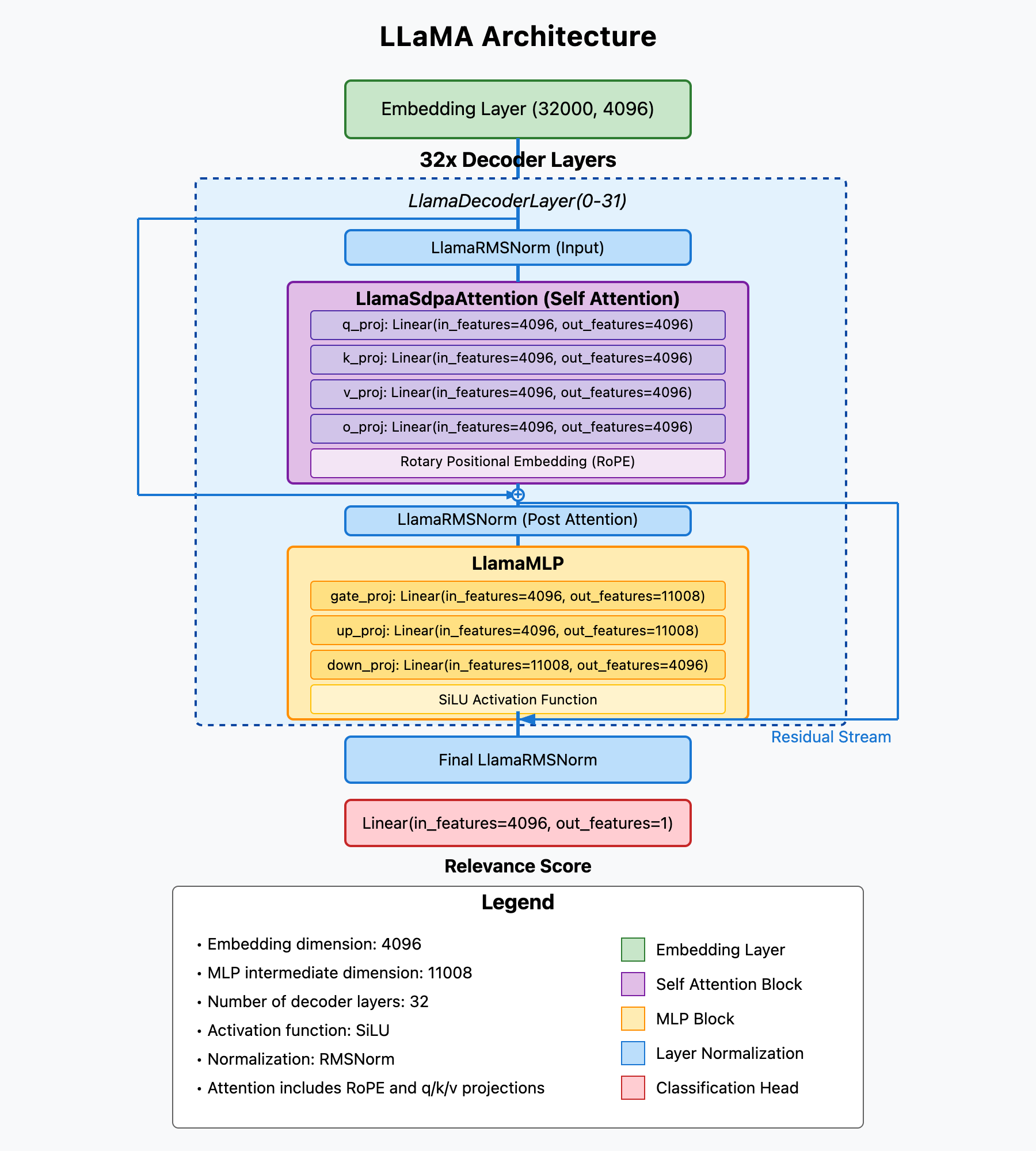} 
    \caption{The RankLlama2-7B internal architecture with 32 layers and 4096 dimensional vectors. Similarly, the 13B models contains 40 layers and 5120 dimensional vectors. Note the score layer, added to the general Llama2 architecture during fine-tuning using LoRa. }
    \vspace{-7mm}
    \label{fig:rankllama-arch}
\end{figure}
\begin{table*}[ht]
\centering
\caption{The Ranking effectiveness of various fine-tuned ranking LLMs, in comparision to the performance of a BM25 reranker. We note that while LLMs are effecient for reranking, they are opaque and complex to understand.}
\begin{tabular}{llcccc}
\toprule
\textbf{Model Name} & \textbf{Base Model}  & DEV MRR@10 & DL19 NDCG@10 & DL20 NDCG@10 \\
\midrule
BM25 & Lucene~\cite{katsimpras2024genra} & 16.7 & 50.6 & 48.0\\
\hdashline
RankLLama2-7b  & Llama2~\cite{touvron2023llama}  & 44.9 & 75.6 & 77.4 \\
RankLLama2-13b & Llama2~\cite{touvron2023llama}  & 45.2 & 76.0 & 77.9 \\
RankLlama3-8b & Llama3~\cite{dubey2024llama}   &   42.8   &  74.3    &   72.4    \\
RankPythia-6.9b & Pythia~\cite{biderman2023pythia}  &  42.9   &   76.0   &   71.5    \\

\bottomrule
\end{tabular}

\label{tab:ranking_performance}
\end{table*}

\subsection{Inner Interpretability in IR}
Most works of interpretability in information retrieval ~\cite{anand2022explainable} have been model extrinsic~\cite{rahimi2021explaining,singh2019exs,chowdhury2023rank}. Among model \textit{intrinsic} interpretability methods, some works focus on using gradient-based methods to identify important neurons ~\cite{chen2024axiomatic,fernando2019study}. Although gradient-based methods give an accurate perspective on the flow of information, they are too fine-grained to give a human-level understanding of the LLM's inner circuit~\cite{anand2022explainable}. A number of works have attempted to examine the inner workings of neural retrievers to understand if they satisfy IR axioms and/or to spot known features. Fan et al. \cite{fan2021linguistic} probe fine-tuned BERT models on three different tasks, namely document retrieval, answer retrieval, and passage retrieval, to find out if these different forms of relevance lead to different models. Zhan et al. ~\cite{zhan2020analysis} study the attention patterns of BERT after fine-tuning
on the document ranking task and make note of how BERT dumps duplicate attention weights on high frequency tokens (like periods). Similar to them, Choi et al.~\cite{choi2022finding} study attention maps in BERT and report the discovery of the IDF feature within BERT attention patterns. ColBERT investigations~\cite{formal2021white,formal2022match} study its term-matching mechanism and conclude that it captures a notion of term importance, which is enhanced by fine-tuning. MacAvaney et al. ~\cite{macavaney2022abnirml} propose ABNIRML, a set of diagnostic probes for neural retrieval models that allows searching for features like writing styles, factuality, sensitivity and word order in models like ColBERT and T5.  Parry et al. ~\cite{parry2024mechir} provide a framework for diagnostic analysis and intervention within Neural Ranking models by generating perturbed inputs followed by activation patching. 

These studies, however, have been conducted in the pre-LLM era and hence are much smaller in size. It is unknown if the findings from BERT will carry over to LLMs like Llama, given that BERT is an encoder-only model whereas most modern LLMs are decoder-only models. While most of the above works focus on attention heads and pattern, our work focuses on probing the MLP activation layers, which are now believed to be the primary feature extraction location within the LLM \cite{gurnee2023finding}. Parallel to our work, Liu et al.~\cite{liu2025large}, through activation patching experiments, identify a progressive, multi-stage process where LLMs extract document and query information in early layers, process relevance according to instructions in middle layers, and use specific attention heads in later layers to generate relevance judgments. Recently Nijasure et al.~\cite{nijasure2025relevance} conduct a behavioral exploration of LoRA fine-tuned LLMs for Passage Reranking to understand how relevance signals are learned and deployed by them. To the best of our knowledge, our work is one of the first few works towards the study of neuron activations in LLMs for a large scale of statistical and semantic features.

\section{Identifying  Context Neurons}
We describe our probing pipeline to identify context neurons - neurons that are sensitive to or encode desired features - in the LLM architecture. 

\textbf{Ranking Models:}
For our Ranking LLM interpretability study, we first select RankLlama \cite{ma2023fine}, an open source pointwise reranker that has been fine-tuned on Llama-2  with the MS MARCO dataset using LoRA. Given an input sequence (query Q, document set D), RankLlama reranks them as :
\begin{align*}
input &= \mathrm{'}query: \{Q\} \ document: \{D\} </s>\mathrm{'}   \\
Sim(Q,D) &= Linear(Decoder(input)[-1])
\end{align*}
where Linear($\cdot$) is a linear projection layer that
projects the last layer representation of the end-of sequence token to a scalar. The model is fine-tuned using a contrastive loss function. RankLlama is accessible on Huggingface\footnote{\url{https://huggingface.co/castorini/rankllama-v1-7b-lora-passage}} and has demonstrated near state-of-the-art performance on the MS MARCO Dev set, as noted by Ma et al.~\cite{ma2023fine}. We experiment on both the 7b and 13b parameter models fine-tuned for passage reranking. 

To facilitate comparison with other ranking architectures, we additionally fine-tune Llama3.1-8b and Pythia-6.9b using an approach similar to the fine-tuning methodology employed by the Rank\-Llama authors. Each of these LLMs is fine-tuned with a LoRa rank of 32 for 1 epoch, and their performance is evaluated on the commonly used TREC DL19 and DL20 datasets. Table \ref{tab:ranking_performance} compares the ranking performance of all LLMs used in this study, indicating that, once fine-tuned, they achieve similar effectiveness.

\textbf{Internal Architecture:}
We refer to the internal architecture of RankLlama2-7b in Figure \ref{fig:rankllama-arch}. Each block of the Llama transformer architecture can be divided into two main groups: the Multi-Head self attention layers and the MLP layers. For example, the Llama-2 7b and 13b architectures  consist of 32 and 40 such identical blocks, where each component has a dimensionality of 4096 and 5120 respectively. The feed-forward sublayer takes in the output of the multi-head attention layer and performs two linear transformations over it, with an element-wise non-linear function $\sigma$ in between.
\begin{align*}
    f_i^l = \mathbf{W}_v^l \times \sigma(\mathbf{W}_k^la_i^l+b_k^l) + b_v^l
\end{align*}
where $a_i^l$ is the output of the MHA block in the $l$ layer, and $\mathbf{W}_v^l$, $\mathbf{W}_k^l$, $b_k^l$, and $b_v^l$ are learned weight and bias matrices.

This MLP layer applies pointwise nonlinearities to each token independently and therefore performs the majority of feature extraction for the LLM. Additionally, the MLP layers account for $\frac{2}{3}$rd of total parameters. As a result, our study focuses on the value of the residual stream (also known as hidden state) for token $t$ at layer $l$, right after applying the MLP layers. Elhage et al.~\cite{elhage2021mathematical} provide further details about the transformer architecture math.

\begin{figure*}[t]
    \centering
    \includegraphics[width=\linewidth]{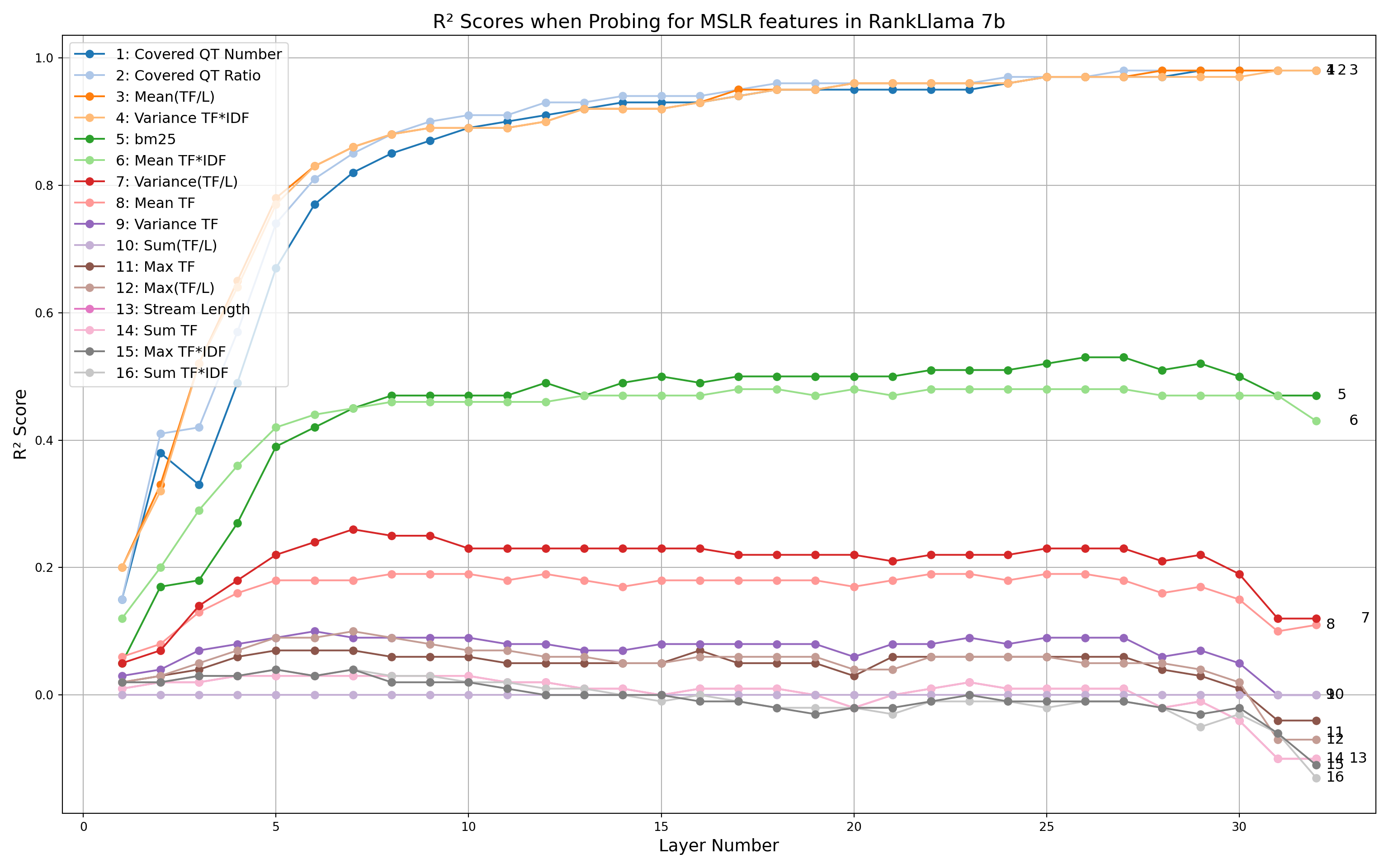}
    \caption{Probing for statistical features from the MSLR dataset in RankLlama2-7b model. Here $QT$ stands for Query Term, $TF$ stands for Term Frequency and $\cdot/L$ stands for length normalized. The graph lines indicate the presence of a particular feature along the layers of the LLM.  Other features like $Covered \ QT \ Number$, $Covered \ QT \ Ratio$, $Mean(TF/L)$ and $Variance \ TF*IDF$ show increasing prominence from the 1st layer to the last, ultimately playing an important role in ranking decision making. Other MSLR features like $Sum(TF/L)$, $Max(TF/L)$, and $Sum \ TF*IDF$ show negative correlation with RankLlama decision making. }
    \label{fig:ep1}
\end{figure*}

\textbf{Activation Sourcing:}
We set up PyTorch transformer forward hooks to mine activations from the ranking LLMs. We tokenize and feed the query-document pairs to the pointwise LLM discussed above, and capture activations corresponding to each input sequence across all layers. The dimension of each layer's activation is the number of tokens times the number of hidden units in the LLM. This dimension is 4096 in Llama2-7b and 5120 in Llama2-13b. To reduce complexity of the computation, we aggregate activations across tokens in an input sequence within a layer. Following Gurnee et al. ~\cite{gurnee2023finding}, we try out mean and max activation aggregation to ultimately obtain a $4096/5120$ dimensional activation vector per layer. This corresponds to aggregated activations for $4096/5120$ neurons in each layer, which is the focal point of our study. We quantize and store these activation tensors to be used later for further experiments. Intuitively, we have captured the internals of the model at this point, in which we can start searching for desired features.

\begin{figure*}
    \centering
    \includegraphics[width=\linewidth]{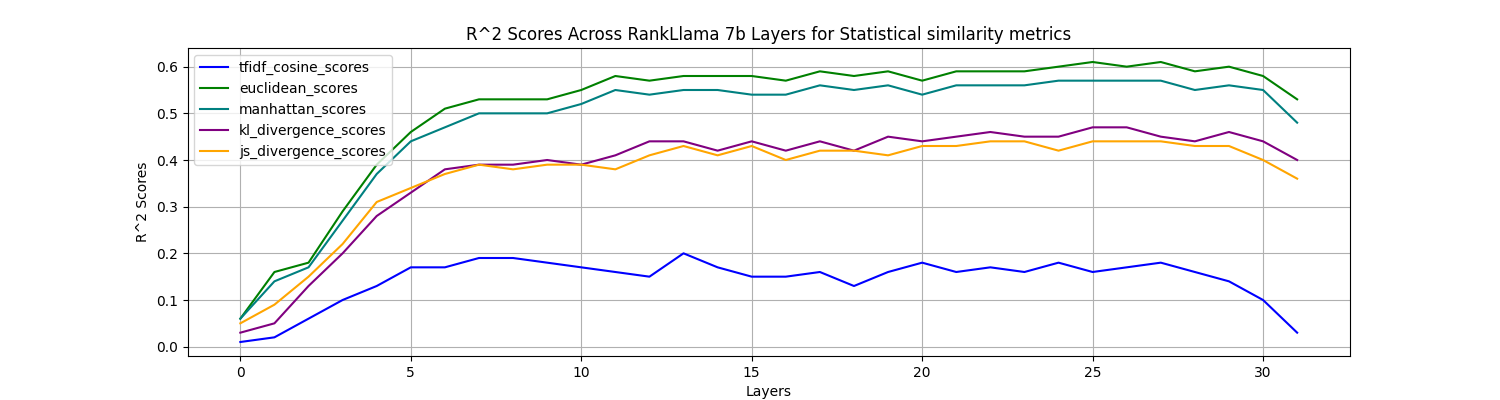}
    \caption{Plot showing $R^2$ scores of statistical query-document distance metrics when used to probe Rankllama2-7b. Scores indicate that these distance metrics are not encapsulated within Rankllama2-7b as is.  }
    \label{fig:ep1c}
\end{figure*}

\textbf{Target Features:}
The MSLR dataset~\cite{DBLP:journals/corr/QinL13}  provides a collection of features that that can be used for learning to rank experiments. Because we are calculating these features ourselves for some additional documents, many of the MSLR features are not usable in our study -- e.g., we do not have anchors, links, URLs, or similar information. Consequently, we aim to search for a subset of these MSLR features within the LLM activations. Specifically, we focus on mining the following 19 features from the MSLR dataset within the LLM (``stream'' means the document or passage text): \emph{Min TF, Min(TF/L), Min TF*IDF, Covered QT Number, Covered QT Ratio, Mean(TF/L), Variance TF*IDF, BM25, Mean TF*IDF, Variance(TF/L), Mean TF, Variance TF, Sum TF, Max TF, Max(TF/L),  Stream length, Sum TF, Max TF*IDF and Sum TF*IDF}. We do not claim that this is a definitive list of features universally acknowledged as important for ranking; they are chosen because they are well known.


In addition to MSLR features, we mine for known query-document similarity metrics within the LLM architecture. For this we consider five traditional query-document similarity scores like \emph{tf*idf cosine scores, Euclidian score, Manhattan score, KL-divergence score, Jensen-Shannon divergence score} as well as two popular semantic relevance metrics like \emph{BERT} and \emph{T5} scores. 

Each feature is probed individually: for instance, the probe for stream length is dedicated solely to that feature without searching for any others. Although many of these features are known to be correlated and may share neurons, this study does not include correlation analysis among features.

\textbf{Probing Datasets:}
To facilitate probing~\cite{sajjad2022neuron}, we need to curate a dataset of input sequences to study the model. We select query-document pairs from the MS MARCO test set \cite{nguyen2016ms} for this purpose. For each query, we include documents that are highly relevant, highly non-relevant, and of intermediate relevance. Our input sequences for the ranking LLM consist of these query-document pairs. Note that it is important for each probing dataset to be balanced -- i.e., contain a uniform number of samples across the range of values of the feature being probed \cite{gurnee2023finding, belinkov2022probing}. For instance, when analyzing the BM25 feature, our dataset included a balanced ratio of query-documents with both low and high BM25 values. Having an unbalanced probing dataset can lead to a biased analysis as the model might have a tendency to overfit the dominant class. For each input sequence, we calculate the expected value of the feature being studied. Consequently, we compute the values for each of the 19 MSLR features and 7 similarity scores for our query-document pairs.

\textbf{Identifying context neurons:}
After collecting activations and feature labels for a wide range of input sequences (query-document pairs), we start the process of identifying context neurons. Unlike most prior research that employs classification-based probes \cite{macavaney2022abnirml,fan2021linguistic}, we utilize ridge regression-based probes to capture the continuous nature of the features under study. We first split the activations and their corresponding labels into training, validate and testing sets (60:20:20). 

Subsequently, we perform a layer-wise analysis. For each layer and feature, we employ a sparse probe by fitting the activation vectors of the training split to the corresponding feature labels. To maintain the sparsity of the context neurons, we use Lasso (L1 regularization) with $\alpha=0.1$. We perform 5-fold cross validation to avoid overfitting and provide a more reliable estimate. After fitting the activations to the feature's labels, we compute the $R^2$ score to measure how well the regression curve explains the variability of the labels. $R^2$ ranges from 0 to 1,  with 1 indicating that the curve perfectly explains the varibiality in the data. A high \( R^2 \) score indicates the presence of a neuron that is sensitive to or activates for the feature being studied, whereas a negative \( R^2 \) confirms a negative correlation between the feature and the neuron's activations.

\textbf{Experimental Details:}
For our probing experiments, we select 500 queries from the MS MARCO dev set and retrieve the top 100 documents for each query from the MS MARCO collections corpus using a BM25 ranker. We then compute activations corresponding to each query-document pair using all four models. The results reported in the following section are after mean aggregation and quantization for efficient storage. The size of each probing dataset is greater than 5,000 to ensure incorporation of a wide range of values for the feature being probed. The 100 documents retrieved for a query serve as the query's corpus for idf calculation. We use the bert-base and t5-base models for BERT and T5 score computations. We use the well-known Okapi implementation of BM25 in our experiments.

\begin{figure*}[t]
    \centering
    \includegraphics[width=0.6\linewidth]{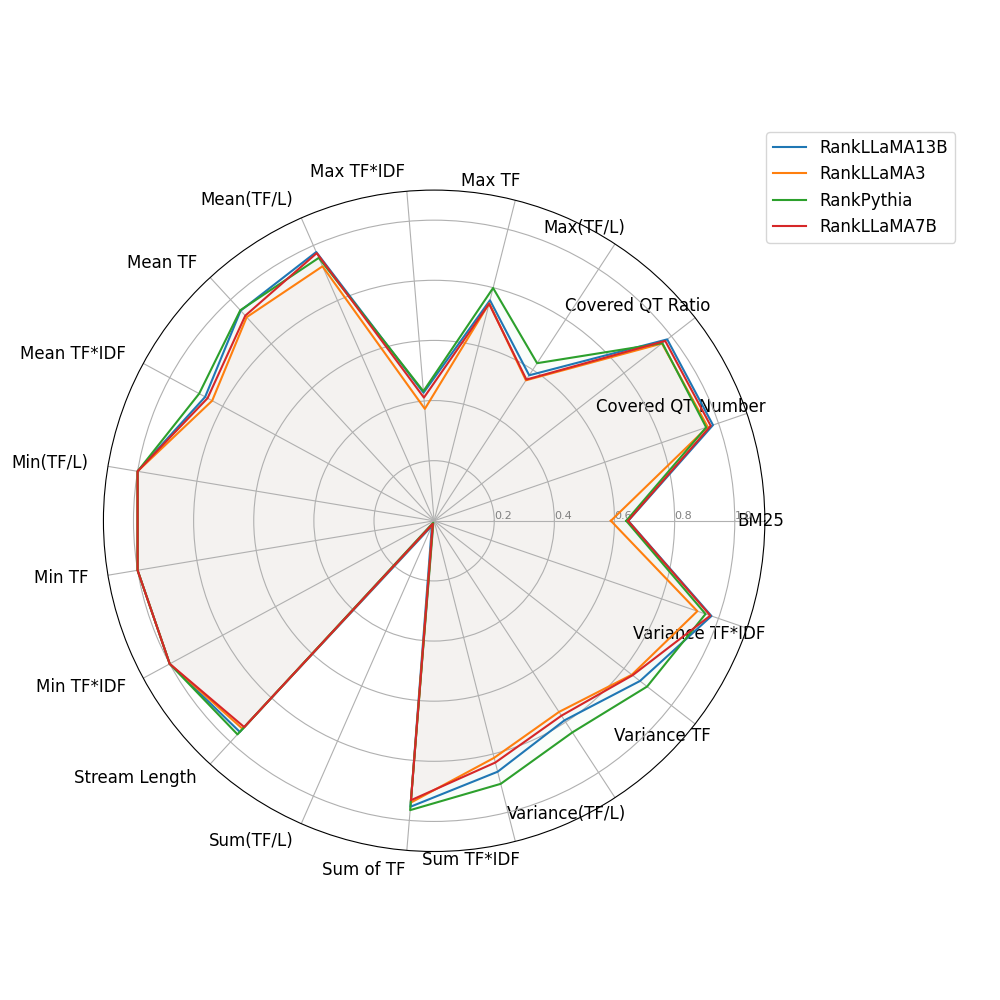}
    \vspace{-1.2cm}
    \caption{Plot comparing the maximum $R^2$ score of each query-document IR feature metrics when used to probe different Ranking LLM architectures - RankLlama2-7b and 13b, RankLlama3-8b and RankPythia-6.9b. Here $\cdot/L$ stands for length normalized and $QT$ stands for Query Term. We observe that different LLM architectures, when fine-tuned on the same dataset and loss function, learn very similar ranking related latent features. This suggests, that our probing findings aren't LLM specific, but generalizable. }
    \label{fig:spider_graph}
\end{figure*}

\section{Research Questions}

\subsection{Statistical features within RankLlama2-7b}
\subsubsection{MSLR features} For the first set of experiments, we designed sparse probes for each of the selected MSLR features and conducted experiments on each layer of the RankLlama2-7b architecture. All experiments were run with both mean and max activation aggregation over the input sequence. Our findings (Figure \ref{fig:ep1}) using mean aggregation and are categorized into two groups: (i) Features that exhibit a strong fit in certain layers, (ii) Features that do not correlate to the LLM activations in any layer.

\textbf{Positives}: Features that achieve an \( R^2 \) score greater than 0.85 in any particular layer are considered strong positives. This means it is highly likely that the activation has extracted the particular feature within the neuron. We found that four metrics, namely \textit{covered QT Number}, \textit{covered QT Ratio}, \textit{Mean(TF/L)}, and \textit{variance of TF*IDF} frequently exhibit low Mean Square Errors and high \( R^2 \) scores. This indicates that there exist MLP neurons within the investigated layers that perform feature extractions similar to these MSLR features. Additionally, we observed that the \( R^2 \) scores obtained with mean and max activation aggregation are comparable, except minor exceptions, suggesting either aggregation method is suitable.

\textbf{ Negatives}: Certain features failed to achieve an \( R^2 \) score greater than 0.1 in the final layer, often achieving a highly negative \( R^2 \). This indicates that the LLM does not consider these features important in their current form. The 10 features from the MSLR set that fell into this category include: \emph{Sum and Max of TF*IDF}, \emph{Sum(TF/L) and Max(TF/L)}, \emph{Max and variance of TF}, \emph{stream length} and \emph{Sum of TF}. Thus, while \emph{Mean(TF/L)} is something the LLM highly seeks, it does not seek for the \emph{sum} and \emph{max} of \emph{normalized term frequency} at all.


\subsubsection{Traditional similarity metrics}
We also probe the LLM architecture for non-semantic statistical query-document distance metrics like \emph{tf*idf cosine score}, \emph{euclidean score}, \emph{manhattan score}, \emph{Kullback-Leibler divergence score} and  \emph{Jensen-Shannon divergence score}. Our motive behind this is to identify if the LLMs include components that mimic statistical similarity measures. Results for the probe on RankLlama2-7b are visualized in  Figure \ref{fig:ep1c}. 
We observe that statistical score based methods do not correlate well with neuron activations. Out of the feature examined, \emph{euclidean score} shows the highest correlation with $R^2$ reaching $0.6$,  dipping before the final layer.  Given its well known success in the past, it is ironic that the \emph{tfi*df cosine score} between the query and document shows the least correlation when probed for, within RankLlama activations.

\begin{figure*}
    \centering
    \includegraphics[width=0.9\linewidth]{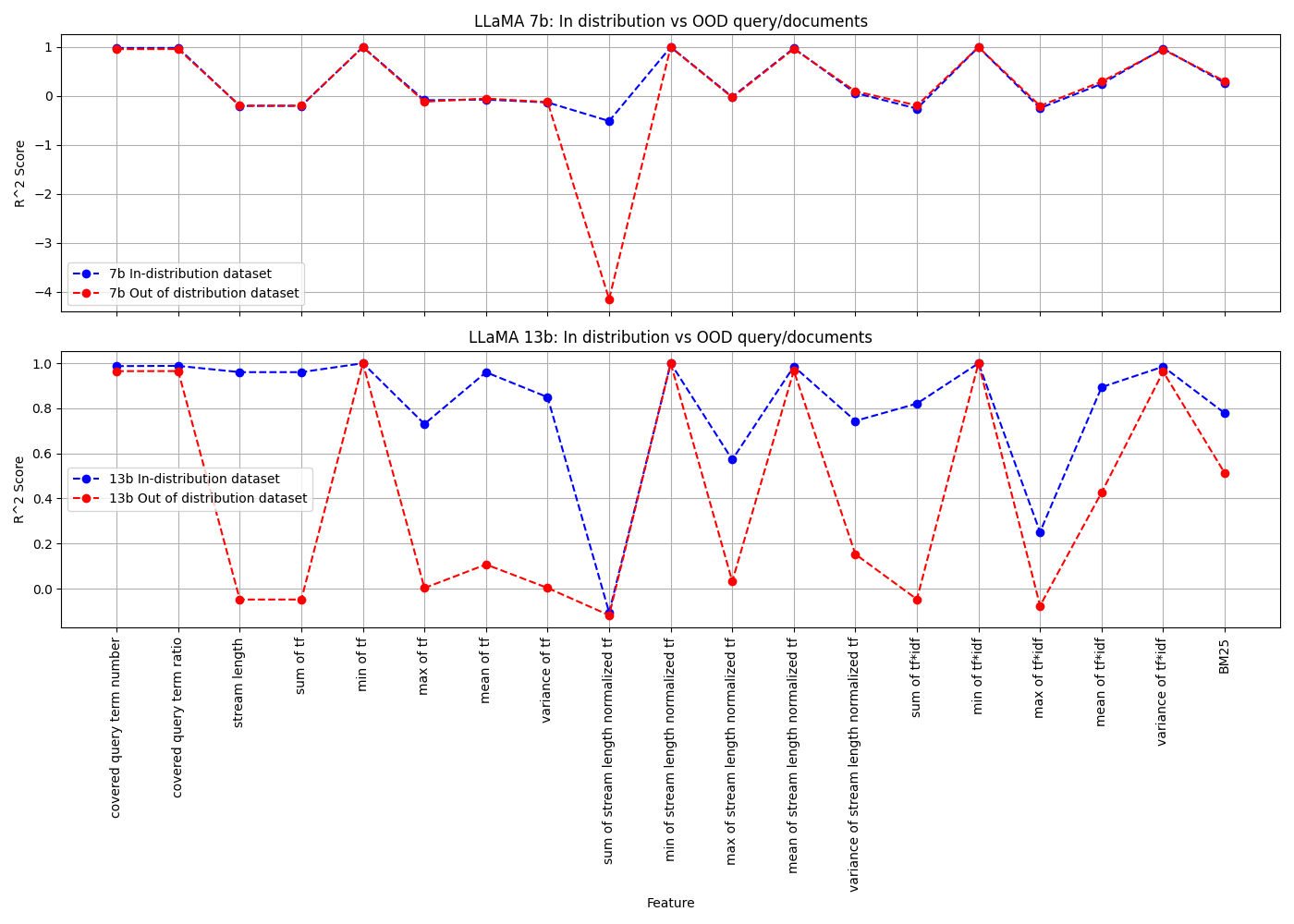}
    \caption{Probing the final layers of RankLlama2-7b and 13b with in-distribution vs out-of-distribution datasets. We witness that most MSLR features when probed for in RankLlama2-7b, show similar performance with both in-distribution and out-of-distribution datasets. This is however not the case with RankLlama2-13b, where certain features like \textit{stream length} and \textit{sum of term frequency} show a strong presence in the in-distribution dataset probe, even though they are unlikely features to  influence a ranking decision. This suggests overfitting on the MS MARCO dataset in RankLlama2-13b.}
    \label{fig:ep4}
\end{figure*}

\subsubsection{BERT and T5 scores}
BERT and T5 are neural models widely used for retrieval and reranking tasks before the advent of LLMs. LLMs are much larger in size compared to BERT models. As a result it is of interest to see if BERT and T5 subnetworks are present within the studied LLM's activations. We design probes to mine for the cosine distance between the BERT and T5 embeddings of the query and the document. Our observations show that both BERT and T5 obtain moderate $R^2$ scores in our probes, reaching $0.7$ and $0.82$ on RankLlama2-7b and 13b respectively. The BERT and T5 scores follow each other across the LLM layers. This suggests the LLM likely does not encode BERT and T5 subnetworks as-is. We compare our findings to other work on BERT probing in Section \ref{section:Discussion}.

\subsection{Comparing Ranking LLMs}
We next explore whether different ranking LLMs encode similar or distinct Information Retrieval features. To investigate this, we extend our analysis beyond the RankLlama2-7b model and compare it to our two additional fine-tuned LLMs, RankLlama3-8b and RankPythia-6.9b, as well as the RankLlama2-13b model. We fine-tune these models under an identical passage reranking framework, with an identical dataset and loss function, to ensure a fair and consistent comparison.

Once the models are fine-tuned, we conduct the same probing experiments to analyze how well each ranking LLM encodes the known human-engineered and semantically meaningful attributes relevant to ranking tasks. The results of these probing experiments are summarized and visualized in a spider chart (Fig. \ref{fig:spider_graph}), allowing for an intuitive comparison of feature encoding across the four models. It highlights the similarities and differences in the latent features captured by LLMs with varying architectures and sizes, shedding light on the generalizability and architecture-specific tendencies of ranking-focused fine-tuning. The figure depicts the $R^2$ score for an IR feature, when probed in a particular LLM, and lies between $[0,1]$.

All four ranking LLMs demonstrate broadly comparable patterns for most features. They appear closely aligned on coverage-based features---such as Covered Query Term Number and Ratio---with each scoring in the high 0.95--0.98 range. By contrast, the greatest differences emerge in TF-based statistics like \emph{Max of Term Frequency} and \emph{Variance of TF/TF*IDF}, where RankPythia tends to exhibit slightly higher peaks (e.g., a larger $R^2$ for \emph{Max of TF}) than the RankLLaMA models. Additionally, all LLMs show near zero performance in encoding \emph{Sum of Stream Length Normalized TF}, suggesting similarity in how these models handle term-frequency weighting. Overall, the four models encode IR features in a largely similar manner, indicating that our probing findings are generalizable and not LLM-specific.

\begin{figure*}
    \centering
    \includegraphics[width= \linewidth]{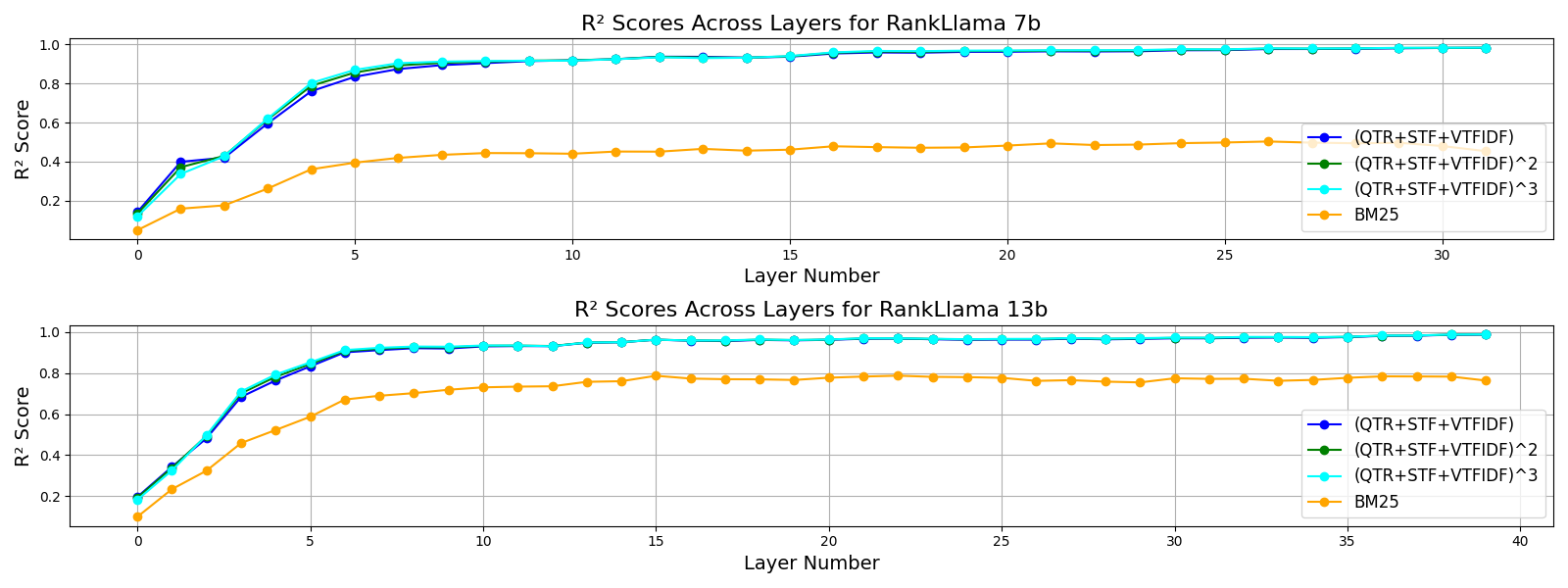}
    \caption{Graph showing $R^2$ scores of  feature groups $(QTR+STF+VTFIDF)$, $(QTR+STF+VTFIDF)^2$ and $(QTR+STF+VTFIDF)^3$ over the layers of the RankLlama2-7b and 13b architectures, where $QTR$ represents \textit{covered query term ratio}, $STF$ represents \textit{mean stream length normalized term frequency} and $VTFIDF$ represents \textit{variance of tf*idf}. }
    \label{fig:ep2}
\end{figure*}

\subsection{Out-of-distribution queries and documents}
All four of the LLMs we study were obtained by fine-tuning the base models using LoRa on the MS MARCO dev set \cite{ma2023fine}. It is of interest to see if during inference the fine-tuned LLM extracts different features from in-distribution and out-of-distribution query-document pairs. We try probing all the studied LLMs with query-document pairs from two others datasets. First, we use the BEIR Scidocs dataset \cite{thakur2021beir}, comprising scientific documents and queries. We select 200 queries and use bm25 to retrieve the top 100 documents for each. We then probe the MS MARCO fine-tuned LLMs with BEIR activations. We repeat the process with the SoDup dataset \cite{h4stackexchange}, which contains a question from StackOverflow as query and a list of other relevant questions from StackOverflow as potential duplicate candidates. 
For each question, we select known duplicate questions from the dataset and treat them as relevant documents. We again pick 200 queries from this dataset for our out-of-distribution probing experiments. We probe for each of the 19 MSLR features on all the fine-tuned ranking LLMs and compare probing results between in-distribution and the out-of-distribution datasets. We show our comparision of the probing results of RankLlama2-7b vs RankLlama2-13b in Figure \ref{fig:ep4}, reporting the $R^2$ score of each feature in the last layer of the respective LLM. For the feature \emph{mean of tf} we see that our probes cannot find any activations representing the feature in RankLlama 7b, both with in-distribution and out-of-distribution datasets. However, when probing RankLlama 13b, while out-of-distribution activations do not capture this feature, it gets a hit within in-distribution activations.

We see some variance in results between RankLlama2-7b and RankLlama2-13b in this set of experiments. The probes seem to fare similarly on most features between in-distribution and out-of-distribution data on RankLlama2-7b. The only exception being \emph{sum of stream length normalized tf}, which has a negative $R^2$ score for both in and out distribution data, making its magnitude insignificant. However, the probes seem to fetch different results between in and out-of-distribution data for RankLlama2-13b. It in particular finds probes strongly correlated to \emph{stream length}, \emph{sum} and \emph{mean of tf}, and \emph{mean of tf*idf} in the in-distribution data, but \textit{not} for the out-of-distribution data. A likely reason for this might be that RankLlama 13b has overfit to the MS MARCO dev set and is hence seeking features like \emph{stream length} which are known to not generalize for reflecting query-document relevance \cite{belinkov2022probing}.

We compared the probing performance of in-distribution and out-of-distribution query-document pairs in RankLlama3-8b and RankPythia-6.9b, observing similar performance for both types of inputs, consistent with the results observed for RankLlama2-7b in Figure \ref{fig:ep4}. The graphs depicting these probing results are not included due to space constraints.

\subsection{Feature Groups within RankLlama}
In the previous subsection, we found that several MSLR features (\textit{covered QT Number, covered QT Ratio, mean(TF/L)} and \textit{variance of TF*IDF}) appear to be modeled within LLM activations, especially within the later layers. These features also seem to be a fair choice to model relevance based on intuition. However, these features might not be present individually but in combination with one another or individually with different exponents. To test for this possibility, we probe for combinations of these features within different layers of the LLM and find various combinations of  \textit{ covered query term ratio, mean of stream length normalized term frequency} and \textit{variance of tf*idf}. We find strong indication for their combined presence within the LLM activations of the later layers of the LLM. For example, if QTR represents \textit{covered query term ratio}, STF represents \textit{mean of stream length normalized term frequency} and VTFIDF represents \textit{variance of tf*idf}, we find high scores on average for all of QTR+STF, QTR+VTFIDF, STF+VTFIDF, QTR+STF+VTFIDF, QTR*STF, STF*VTFIDF, QTR*VTFIDF, and QTR*STF*VTFIDF when probing the last layer of the LLM. This potentially indicates that the LLM has over the course of layers learned some representation of (QTR+STF+VTFIDF)$^k$. In Figure \ref{fig:ep2}, we show the probing performance of a sum of those three features and their exponents, relative to BM25 in RankLlama2-7b and 13b models. We observe that the sum of this feature group, its square, as well as its cube consistently show a strong correlation to neuron activations within RankLlama. This analysis provides a foundation for future work to uncover and investigate complex groups of features encoded within LLM activations. By highlighting how various ranking architectures capture and represent different IR features, it opens avenues for deeper exploration into latent feature interactions and the mechanistic underpinnings of ranking-focused language models.



\balance
\section{Discussion} \label{section:Discussion}

\textbf{Validating Probing Results}
While probing techniques are widely used to understand the internal workings of LLMs, there are several limitations of using probing for this purpose~\cite{belinkov2022probing}. (1) Probing techniques reveal correlations between specific features and neuron activations, but not a causal analysis of the decision making process. (2) The insights gained from probing are heavily dependent on the dataset used for probing. If the dataset is not representative of the model's typical input, the results may not accurately reflect the model's general behavior. (3) LLMs often rely on complex interactions between multiple features. Probing techniques may fail to capture these interactions, leading to an incomplete understanding of how the model makes decisions. (4) Probing can identify features but may mislead regarding their role or significance in the model's decision-making process. However, many of the limitations of probing can be addressed by employing various techniques, such as creating balanced probing datasets and independently validating probing results through methods like ablation studies and feature attribution analyses. 

To partially validate our probing results, we analyzed neurons in the final layer of RankLLaMA models that exhibited strong correlations ($R^2 > 0.85$) with probed features, as this layer directly influences ranking decisions. Approximating Shapley values ~\cite{datta2016algorithmic,lundberg2017unified,chowdhury2024rankshap} for feature attributions via Integrated Gradients~\cite{sundararajan2017axiomatic}, we assessed the contributions of individual neurons and neuron groups to the ranking predictions, leveraging a value function based on changes in NDCG scores for MS MARCO query-document pairs. Probing validation was conducted on 100 query-document pairs for each RankLLaMA model, and we computed average attributions across neuron groups of varying sizes. Results showed that identified neuron groups ranked within the 95th percentile in 79 out of 100 cases for RankLLaMA 7B and 84 out of 100 for RankLLaMA 13B, confirming that these neurons are instrumental in ranking decisions and validating the location of the probes. 

\textbf{Comparing Notes with Previous Probing Work:}
A number of studies have probed BERT and T5 with the aim of understanding if they encode concepts like term frequency and inverse document frequency. Formal et al.~\cite{formal2021white} study the matching process of ColBERT \cite{khattab2020colbert} and conclude that the model is able to capture a notion of term importance and relies on exact matches for important terms. This is in agreement with our findings, where we find strong correlations to term matching features like \emph{covered query term number/ratio} and \emph{stream length normalized term frequency} within LLM activations. In a later work, Formal et al.~\cite{formal2022match} conduct another study to measure the out-of-domain zero-shot capabilities of BERT/T5 models in lexical matching on the BEIR dataset,  and show that these models fail to generalize lexical matching in out-of-domain datasets or terms not seen at training time. This finding continues for RankLlama as well, where the model is unable to generalize lexical matching to terms not seen beforehand (Figure \ref{fig:ep4}, RankLlama 13b OOD vs In-distribution probing discrepencies).

\section{Conclusion and Future Work}
In this study, we probed four different large language models, each fine-tuned for passage reranking, to look for a subset of features from the MSLR dataset. These human-engineered features, which include query-only, document-only, and query-document interaction features, are recognized as significant in ranking tasks. Using a layer-wise probe, we discovered that the activations in most layers could accurately replicate select features such as \emph{covered QT Number/Ratio, Var(TF/IDF)}  and \emph{Mean(TF/L)}. This suggests that the fine-tuned Llama network deems these features important. Conversely, we found no correspondence for features like \emph{sum} and \emph{max} of \emph{tf*idf} and \emph{sum} and \emph{max} of \emph{tf} in the probed neurons. This indicates the absence of these features in RankLlama's neural feature extractors. We also reported observations on feature \textit{groups} that show a strong correlation with RankLlama neurons and hypothesized that some abstract features are mined by the LLM. We also observed that different LLM architectures, when fine-tuned on the same dataset with the same loss function, exhibit similar latent features when probed. Finally, we compared later features encoded between in-distribution and out-of-distribution datasets and encountered a case of the LLM's apparently overfitting during fine-tuning. These findings enhanced our understanding of ranking LLMs and gave us generalizable insights for the design of more effective and transparent ranking models.
The long-term objectives of this endeavor include: (i) identifying potential modifications to existing MSLR features, such that they further align with LLM activations, (ii) deciphering and reverse-engineering segments of the LLM that do not correspond to recognized MSLR features, and (iii) ultimately cataloging all features deemed significant by LLMs and plugging them back into simpler statistical models to enhance the performance and interpretability of statistical ranking models. These directions will help further explain the underlying mechanisms of ranking LLMs and improve human trust in them.

\subsection*{Acknowledgments}
This work was supported in part by the Center for Intelligent Information Retrieval. Any opinions, findings and conclusions
or recommendations expressed in this material are those of the authors and do not necessarily
reflect those of the sponsor.

\bibliographystyle{ACM-Reference-Format}
\bibliography{sample-base}

\pagebreak
\newpage


\onecolumn

\end{document}